\documentclass[aps,twocolumn,amsmath,amssymb,preprintnumbers,superscriptaddress,floatfix]{revtex4}

\usepackage{physics}

\usepackage[utf8]{inputenc}
\usepackage{newtxtext}
\usepackage[upint]{newtxmath}
\usepackage{microtype}
\usepackage{textcomp}
\usepackage{dsfont}
\usepackage{eucal}
\usepackage{siunitx}
\usepackage{soul}

\usepackage{enumerate}
\usepackage{amsfonts}
\usepackage{color}
\usepackage{soul}

\usepackage{todonotes}
\presetkeys%
    {todonotes}%
    {inline}{}

\usepackage{graphicx}

\usepackage[colorlinks,allcolors=blue]{hyperref}
\usepackage[capitalize]{cleveref} 
\usepackage{cleveref}

\newcommand{\me}[1]{\mathrm{e}^{#1}}                            

\newcommand{\cc}[1]{{#1}^*}                                     

\renewcommand{\v}[1]{\boldsymbol{#1}}                           
\newcommand{\uv}[1]{\v{e}_{#1}}       

\definecolor{DarkBlue}{rgb}{0,0,0.80}
\definecolor{DarkRed}{rgb}{0.80,0,0}
\definecolor{Purple}{rgb}{0.55,0,0.55}
\definecolor{Purple}{rgb}{0,0,0.8}

\newcommand*{\defeq}{\coloneqq}

\newcommand{\up}{\uparrow}                                      
\newcommand{\dn}{\downarrow}                                    

\newcommand*{\TO}{\mathcal T}                     
\newcommand*{\CO}{\mathcal T_c}
\newcommand*{\C}{\mathcal{C}}
\newcommand{\SC}{\textsc{SC}}
\newcommand{\NM}{\textsc{NM}}
\newcommand{\AFI}{\textsc{AFI}}
\newcommand{\intf}{\text{int}}

\let\epsilon\varepsilon

\begin{document}
\title{Spin-pumping in superconductor-antiferromagnetic insulator bilayers}

\author{Eirik Holm Fyhn}
\affiliation{Center for Quantum Spintronics, Department of Physics, Norwegian \\ University of Science and Technology, NO-7491 Trondheim, Norway}
\author{Jacob Linder}
\affiliation{Center for Quantum Spintronics, Department of Physics, Norwegian \\ University of Science and Technology, NO-7491 Trondheim, Norway}

\date{\today}
\begin{abstract}
  We study theoretically spin pumping in bilayers consisting of superconductors and antiferromagnetic insulators.
  We consider both compensated and uncompensated interfaces and include both the regular scattering channel and the Umklapp scattering channel.
  We find that at temperatures close to the critical temperatures and precession frequencies much lower than the gap, the spin-current is enhanced in superconductors as compared to normal metals.
  Otherwise, the spin-current is suppressed.
  The relevant precession frequencies where the spin-current in SC/AFI is enhanced compared to NM/AFI is much lower than the typical resonance frequencies of antiferromagnets, which makes the detection of this effect experimentally challenging.
  A possible solution lies in the shifting of the resonance frequency by a static magnetic field.
\end{abstract}
\maketitle
\section{Introduction}%
\label{sec:introduction}
Both superconductors (SC) and antiferromagnets (AF) are of particular interest in the context of spintronics.
Antiferromagnets disturb neighbouring components less than ferromagnetic or ferrimagnetic materials, because they produce no net stray field~\cite{baltz2018}.
This means that antiferromagnetic components can be packed more tightly and are more robust against external magnetic fields than their ferromagnetic counterparts.
Additionally, antiferromagnets operate at THz frequencies, which are much faster than the GHz frequencies of ferromagnets (F).
This can allow for ultrafast information processing when working with antiferromagnets.

Superconductivity is a type of order that normally competes with magnetism.
However, the discovery of spin-triplet superconductivity has shown that complete synergy between superconductivity and magnetism is possible~\cite{bergeret2005,buzdin2005,linder2015,eschrig2015_rep,linder2019}, and superconductors are now an integral part of spintronics research.
In addition to the potential for minimal Joule heating that comes with superconductivity, superconductors are interesting from a spintronics perspective because of spin-charge separation~\cite{kivelson1990,zhao1995}, which allows spin- and charge-imbalances to decay over different length scales.
It has been observed that the spin relaxation time can be considerably longer than the charge relaxation time~\cite{quay2013}.

Since both superconductors and antiferromagnets are useful as building blocks in spintronic devices, it is of interest to study spin-transport in hybrid superconductor-antiferromagnet devices.
Despite this, SC/AF structures are largely unexplored compared to superconductor-ferromagnetic structures.
Here, we study theoretically spin-pumping in superconductor-antiferromagnetic insulator (SC/AFI) bilayers.
This refers to the injection of a spin-current in the superconductor, which we consider to be spin-singlet and s-wave, by the application of a precessing magnetic field in the AFI~\cite{tserkovnyak2002}.
Spin pumping has been observed in F/SC structures~\cite{jeon2018,jeon2018_nat,yao2018} and investigated theoretically in F/SC structures by calculations based on the local dynamic spin susceptibility in the SC~\cite{inou2017,kato2019} and quasiclassical theory~\cite{silaev2020,silaev2020R}.
The theoretical works found an enhanced spin current in superconductors compared to normal metals (NMs) below the transition temperatures~\cite{inou2017,kato2019}.

While spin-pumping in SC/AF structures has, to our knowledge, not been explored, some important work has been done with normal metal-antiferromagnetic systems.
It has been found theoretically that spin-pumping is of a similar magnitude as in the ferromagnetic case~\cite{cheng2014,kamra2017}, and more recently measurements of the inverse spin-Hall voltage demonstrated the spin-pumping effect in MnF$_2$/Pt~\cite{vaidya2020}.
Combining the demonstration of AF/NM spin-pumping with the above mentioned evidence of F/SC spin-pumping, AF/SC spin-pumping is feasible and merits further study.

We mainly follow the methodology presented in \cite{kato2019}, but modified for a superconductor-antiferromagnetic insulator bilayer.
In particular, the staggered magnetic order of the AFI gives rise to two different scattering channels~\cite{takei2014,fjaerebu2017,fjaerebu2019}, and the two different sublattices can be coupled to the superconductor in a symmetric or asymmetric way.
To capture this we will not approximate the interaction Hamiltonian by a uniform scattering amplitude, as in \cite{kato2019}, but instead model the interaction with an exchange coupling between itinerant electrons in the SC and the localized spins in the AFI.
Using this coupling, it turns out that the relevant quantity is not the \emph{local} dynamic spin susceptibility, as in \cite{inou2017,kato2019}, but instead the \emph{planar} dynamic spin susceptibility.
Using the planar dynamic spin susceptibility we find that the spin-pumping into superconductors from antiferromagnets is enhanced as compared to spin-pumping into normal metals when the temperature is close to the transition temperature and the precession frequency is small compared to the energy gap.
Otherwise the spin-current in the superconductor is suppressed.
This is similar to the results obtained from ferromagnets.
However, unlike in the case of ferromagnets, the resonance frequency in antiferromagnets is typically too large for spin-pumping with frequencies below the gap to be experimentally detectable.
One possible solution is to apply a static magnetic field, which we discuss in \cref{sec:experiment}.

\section{Model}%
\label{sec:methodology}
\begin{figure}[htpb]
  \centering
  \includegraphics[width=0.7\linewidth]{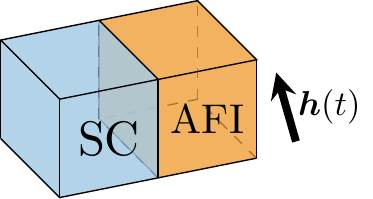}
  \caption{Sketch of a Superconductor (SC)-antiferromagnetic insulator (AFI) bilayer with a precessing external magnetic field $\v h(t)$.}%
\label{fig:sketch}
\end{figure}

The system depicted in \cref{fig:sketch} is modelled by the Hamiltonian 
\begin{equation}
  H = H_\SC + H_\AFI + H_\intf,
\end{equation}
where the Bogoliubov-de Gennes Hamiltonian,
\begin{equation}
  H_\SC = \sum_{\v k \in \square} \mqty(c^\dagger_{\v k, \up} & c_{-\v k, \dn})
  \mqty(\xi_{\v k} & \Delta \\ \cc\Delta & -\xi_{\v k})
  \mqty(c_{\v k, \up} \\ c^\dagger_{-\v k, \dn}),
\end{equation}
where $\square$ is the first Brillouin zone (1BZ) in the superconductor,
gives a mean-field description of superconductivity.
The antiferromagnetic insulator Hamiltonian is given by
\begin{equation}
  H_\AFI = J\sum_{\langle i, j \rangle} \v S_i \cdot \v S_j - K\sum_{i} S_{i, z}^2 - \gamma \sum_i \v S_i \cdot \v h.
  \label{eq:H_afi}
\end{equation}
where $\langle i, j \rangle$ means that the sum goes over nearest neighbours and $\sum_i$ goes over lattice points in the AFI.
The exchange coupling at the interface is given by
\begin{equation}
  H_\intf = -2\sum_i J_i
   \mqty(c^\dagger_{i, \up} & c^\dagger_{i, \dn})
   \v \sigma
  \mqty(c_{i, \up} \\ c_{i, \dn})
  \cdot \v S_i,
\end{equation}
where the sum goes over the lattice points in the interface.
Here, $\xi_{\v k}$, is the kinetic energy measured relative to the chemical potential $\mu$, $c_{\v k,\sigma}$ is the annihilation operator for electrons with spin $\sigma$ and wavevector $\v k$, $J$ is the antiferromagnetic exchange parameter, $K$ is the easy-axis anisotropy, $\v S_i$ is the spin at lattice site $i$ in the AFI and $\gamma$ gives the coupling strength to the external magnetic field $\v h$.
The vector of Pauli matrices is given by $\v \sigma$,  and $J_i = J_A$ ($J_i = J_B$) when $i$ belongs to the $A$ ($B$) sublattice.
Also, $\Delta$ is the superconducting gap parameter, which we assume real and satisfies \begin{equation}
  1 = \lambda \int_0^{\omega_D}\frac{\tanh(\sqrt{\varepsilon^2 + \Delta^2}/2T)}{\sqrt{\varepsilon^2 + \Delta^2}},
  \label{eq:gap}
\end{equation}
where $T$ is the temperature, which we assume to be the same for the superconductor and AFI, and $\omega_D$ and $\lambda$ are material-specific parameters that determine the critical temperature $T_c$ and the zero-temperature gap $\Delta_0 \defeq \Delta(0)$.

In order diagonalize $H_\AFI$ we can do a Holstein-Primakoff transformation followed by a Fourier transform and a Bogoliubov transformation.
This gives to second order in magnon operators the following antiferromagnetic Hamiltonian:
\begin{multline}
  H_\text{AFI} = \sum_{\v k\in\Diamond} \left(\omega^{\alpha}_{\v k} \alpha^\dagger_{\v k}\alpha_{\v k} + \omega^{\beta}_{\v k} \beta^\dagger_{\v k}\beta_{\v k}\right)\\
  + \sqrt{2N_AS} (u_{\v 0} + v_{\v 0}) \gamma\left[h^-\left(\alpha_{\v 0} + \beta^\dagger_{\v 0}\right) + h^+\left(\alpha^\dagger_{\v 0} + \beta_{\v 0}\right)\right],
\end{multline}
where $\Diamond$ is the first magnetic Brillouin zone, which is the 1BZ corresponding to the $A$ sublattice, $N_A$ is the number of lattice points in the $A$ sublattice,
$S$ is the spin at each lattice point,
$\alpha_{\v k} = u_{\v k}a_{\v k} - v_{\v k}b^\dagger_{-\v k}$ and
$\beta_{\v k} = u_{\v k}b_{\v k} - v_{\v k}a^\dagger_{-\v k}$,
where $a_{\v k}$ and $b_{\v k}$ are the magnon annihilation operators for the $A$ and $B$ sublattices,
and
\begin{subequations}
  \label{eq:bogoCoeffs}
\begin{align}
  u_{\v k} &= \frac{Jz + K}{\sqrt{\left(Jz + K\right)^2 - \left(J\gamma_{\v k}\right)^2}},\\
  v_{\v k} &= -\frac{J\gamma_{\v k}}{\sqrt{\left(Jz + K\right)^2 - \left(J\gamma_{\v k}\right)^2}},\\
  \omega^\alpha_{\v k} &= S\sqrt{\left(Jz + K\right)^2 - \left(J\gamma_{\v k}\right)^2} + \gamma h_z, \\
  \omega^\beta_{\v k} &= S\sqrt{\left(Jz + K\right)^2 - \left(J\gamma_{\v k}\right)^2} - \gamma h_z.
\end{align}
\end{subequations}
Here, $h_z$ is the $z$-component of the external magnetic field, which is the same as the magnetization direction in the antiferromagnet and the direction of the easy-axis anisotropy.
Moreover, $h^\pm = h_x \pm i h_y$
and
\begin{align}
  \gamma_{\v k} &= \sum_{\langle \v \delta \rangle} \cos(\v k \cdot \v \delta) = \gamma_{-\v k},
\end{align}
where the sum goes over the nearest neighbour displacement vectors $\v \delta$, and $z$ is the number of nearest neighbours.

To write $H_\intf$ in terms of Fourier components requires us to connect the reciprocal space in the superconductor with the reduced Brillouin zone of the magnetic lattice in the AFI.
This gives rise to so-called Umklapp scattering, where the wavevector falls outside the 1BZ in the AFI~\cite{fjaerebu2019}.
Whether this effect is present depends on the interface.
Depending on how the interface slices the biparte lattice of the AFI, the interface can have a different number of atoms belonging to the $A$ and $B$ lattices.
If the interface has an equal number of atoms from each sublattice and the coupling strengths $J_A$ and $J_B$ are equal, we call it a compensated interface.
Otherwise, it is uncompensated.
We let $\v x = \v 0$ to be the location of a lattice point belonging to the $A$ sublattice and $\v x_0$ be such that all lattice points at the interface can be written $\v x_0 + \tilde{\v x}_i$, where $\v x_0 \cdot \tilde{\v x}_i = 0$.

To capture both compensated and uncompensated interfaces we will use the notation $\delta^A_{\v q_{\parallel},\v k_{\parallel}}=1$ to mean that $\v q \cdot \tilde{\v x}_i - \v k\cdot \tilde{\v x}_i = 2\pi n + d_1$ for all vectors $\tilde{\v x}_i$ such that $\v x_0 + \tilde{\v x}_i$ is in the $A$-sublattice at the interface and for some integer $n$ and a constant $d_1$ that is independent of $\tilde{\v x}_i$.
Similarly, $\delta^B_{\v q_{\parallel},\v k_{\parallel}}=1$ means that $\v q \cdot \tilde{\v x}_i -\v k\cdot \tilde{\v x}_i = 2\pi n + d_2$ for all lattice vectors $\v x_0 + \tilde{\v x}_i$ in the $B$ sublattice at the interface and for some integer $n$ and a constant $d_2$ that is independent of $\tilde{\v x}_i$.
We can determine $d_1$ by noting that both $\tilde{\v x}_i$ and $2\tilde{\v x}_i$ is in the $A$ sublattice, so $2d_1 = d_1 + 2\pi n \implies d_1 = 2\pi m$ for some integer $m$.
Hence, we can set $d_1 = 0$.
Similarly, if $\tilde{\v x}_i$ is in the $B$ sublattice, then $2\tilde{\v x}_i$ is in the $A$ sublattice, so $4\pi n + 2 d_2 = 2\pi m \implies d_2 = l\pi$ for some integer $l$.
The $\v k$-vectors that result in $l$ being an odd-number give rise to the Umklapp scattering channel.
We can drop the superscripts because $\delta^A_{\v q_{\parallel},\v k_{\parallel}}=1 \iff \delta^B_{\v q_{\parallel},\v k_{\parallel}}=1$.
This is because every lattice point in the $B$ sublattice is midway between two lattice points in the $A$ sublattice and vica versa.
Finally, if the number of lattice points at the interface is equal on the superconductor and the antiferromagnet, then half of the possible $\v k$-vectors in the superconductor will give $l = 0$ and the other half will give $l = 1$.
There is a vector $\v G$ connecting the region in $\square$ with $l = 0$ to those with $l=1$.

For a concrete example, consider the situation where the crystal lattices of the SC and AFI are equal and cubical.
The 1BZ in the SC, $\square$, is therefore also cubical.
Meanwhile, the sublattice in the AFI is face-centered cubic, so $\Diamond$ is the truncated octahedron inscribed in $\square$.
A wavevector in the corner of $\square$ will be in the center of the second Brillouin zone in the AFI.
If we let $\v G$ be the vector in a corner of $\square$, then $\exp(i\v G \cdot \v x_i)$ is 1 when $\v x_i$ is in the $A$ sublattice and $-1$ when $\v x_i$ is in the $B$ sublattice.
Thus $\v G$ is the vector that connects the region of $\v k$-vectors in $\square$ with regular scattering and those with Umklapp-scattering.

Using this notation, $H_\intf$ can, to first order in magnon operators, be written
\begin{equation}
  H_\intf = \sum_{\v k \in \square} \sum_{\v q \in \Diamond} \left[T^\alpha_{\v q \v k} \alpha_{\v q} s^-_{\v k} + T^{\beta^\dagger}_{\v q \v k} \beta^\dagger_{\v q} s^-_{\v k}
    + \text{h.c.}
  \right]
  + H^Z_\intf,
  \label{eq:Hint}
\end{equation}
where
\begin{align}
  H^Z_\text{int} &=
  -\sqrt{2SN_A} 
  \sum_{\v k \in \square}
    \delta_{\v k_\parallel,\v 0}\left(\bar J_A - 
    (-1)^l\bar J_B\right)s^z_{\v k}\me{-i\v x_0 \cdot \v k}
    \label{eq:zeeman_term}
\end{align}
is the Zeeman energy and
\begin{subequations}
  \label{eq:transmissionCoeffs}
  \begin{align}
    T^\alpha_{\v q \v k} = -
    \me{i\v x_0 \cdot (\v k + \v q)}\Bigl[ 
    \bar J_A u_{\v q} &+ (-1)^l\bar J_B v_{\v q} \Bigr]\delta_{\v k_\parallel,-\v q_\parallel},
  \\
    T^{\beta^\dagger}_{\v q \v k} = -
  \me{i\v x_0 \cdot (\v k - \v q)}
\Bigl[ 
    \bar J_A v_{\v q} &+ (-1)^l\bar J_B u_{\v q} \Bigr]\delta_{\v k_\parallel,\v q_\parallel} .
  \end{align}
\end{subequations}
Additionally,
\begin{subequations}
  \begin{align}
    \bar J_A = J_A\frac{2\sqrt{2S}N_A^\parallel}{N_S\sqrt{N_A}}, \\
    \bar J_B = J_B\frac{2\sqrt{2S}N_B^\parallel}{N_S\sqrt{N_A}},
  \end{align}
\end{subequations}
where $N_S$ is the number of lattice points in the superconductor and $N_A^\parallel$ ($N_B^\parallel$) is the number of lattice points belonging to the $A$ ($B$) sublattice at the interface,
and
\begin{subequations}
  \begin{align}
    s^z_{\v k} &= \frac 1 2 \sum_{\v q \in \square}
    \left(c^\dagger_{\v q \up}c_{\v q+\v k \up} 
    - c^\dagger_{\v q \dn}c_{\v q+\v k \dn}\right),
    \\
    s^-_{\v k} &=\sum_{\v q \in \square}
    c^\dagger_{\v q \dn}c_{\v q+\v k \up}.
  \end{align}
\end{subequations}
The reason why the factor $(-1)^l$ is in front of the terms proportional to $\bar J_B$ in  \cref{eq:zeeman_term,eq:transmissionCoeffs} is that the coordinate system is defined such that $\v x = 0$ is the location of a lattice point belonging to the $A$ sublattice.

\section{Green's functions}%
\label{sec:green_s_functions}
In order to calculate the spin current we will make use of Green's functions corresponding to three different types of operators.
Let $\psi$ be either $\alpha$, $\beta^\dagger$ or $s^+$, then the lesser, retarded and advanced Green's functions are
\begin{subequations}
  \label{eq:GFs}
  \begin{align}
    G_\psi^<(t_1,t_2, \v k) &= -i\left\langle \psi^\dagger_{\v k}(t_2) \psi_{\v k}(t_1)\right\rangle_0, \\
    G_\psi^R(t_1,t_2, \v k) &= -i\theta(t_1-t_2)\left\langle \left[\psi_{\v k}(t_1),\, \psi^\dagger_{\v k}(t_2)\right]\right\rangle_0,
    \\
    G_\psi^A(t_1,t_2, \v k) &= i\theta(t_2-t_1)\left\langle \left[\psi_{\v k}(t_1),\, \psi^\dagger_{\v k}(t_2)\right]\right\rangle_0,
  \end{align}
\end{subequations}
respectively.
The subscript $0$ means that the expectation values are taken in the absence on $H_\intf$.
This is done because we will treat $H_\intf$ as a perturbation in the interaction picture.
This is a good approximation as long as the the transmission coeffiecients are small and has previously been shown to give good agreement with experiments~\cite{oyanagi2019,kato2020,umeda2018}.
We will also define the distribution function
\begin{equation}
  f^{\psi}(\varepsilon, \v k) \defeq \frac{G_\psi^<(\varepsilon, \v k)}{2i \Im G_\psi^R(\varepsilon, \v k)},
\end{equation}
where the Green's functions in \cref{eq:GFs} are Fourier transformed with respect to the relative time $t_1 - t_2$.
In thermal equilibrium, $f^{\psi}(\varepsilon, \v k)$ is equal to the Bose-Einstein distribution function $n_B(T, \varepsilon)$.

First consider the effect of spin pumping.
We add spin pumping in the AFI by letting $h^\pm(t) = h_0\me{\mp i\Omega t}$.
The reader is referred to \cref{sec:afi_green_s_functions} for the detailed calculation, which shows that the retarded Green's functions are unaffected to second order in $h_0$.
Since the unperturbed Hamiltonian is diagonal in $\alpha$ and $\beta$, this means that the retarded Green's functions for $\alpha$ and $\beta^\dagger$ are
\begin{subequations}
  \begin{align}
    G_{\alpha}^{R}(\varepsilon, \v k) &= \frac{1}{\varepsilon - \omega^\alpha_{\v k} + i\eta^\alpha},\\
    G_{\beta^\dagger}^{R}(\varepsilon, \v k) &= -G_{\beta}^{A}(-\varepsilon, \v k) =  \frac{1}{\varepsilon + \omega^\beta_{\v k} + i\eta^\beta},
  \end{align}
  \label{eq:AFIretarded}
\end{subequations}
where $\eta^\alpha$ and $\eta^\beta$ are the lifetimes of the $\alpha$ and $\beta$ magnons.
The distribution functions are modified by the oscillating magnetic field, and to second order in $h_0$
\begin{multline}
  f^\nu(\varepsilon, \v k) = n_B(\varepsilon, T)\\
  + \frac{2\pi N_AS [(u_{\v 0} + v_{\v 0})\gamma h_0]^2}{\eta^\nu} \delta_{\v k, \v 0} \delta(\varepsilon - \Omega),
  \label{eq:AFIdistribution}
\end{multline}
where $\nu \in \{\alpha,\, \beta^\dagger\}$.

The dynamic spin susceptibility $G_{s^+}^{R}$ is more complicated, but can be calculated from the imaginary time Green's function by use of analytical continuation and Matsubara summation techniques.
This is shown in \cref{sec:bcs_dynamic_spin_susceptibility}, and the result is
\begin{multline}
  G_{s^+}^{R}(\varepsilon, \v k)
  = -\frac 1 4 \sum_{\v q}\sum_{\omega = \pm E}\sum_{\tilde\omega = \pm\tilde E}
  \left(1 + \frac{\xi\tilde\xi + \Delta^2}{\omega\tilde\omega}\right)
  \\ \times
  \frac{n_F(\tilde\omega, T)- n_F(\omega, T)}{\varepsilon + i\eta^\SC - (\tilde\omega - \omega)},
  \label{eq:BCSspinSusc}
\end{multline}
where $\xi = \xi_{\v q}$, $\tilde \xi = \xi_{\v q + \v k}$, $E = \sqrt{\xi^2 + \Delta^2}$ and $\tilde E = \sqrt{\tilde\xi^2 + \Delta^2}$, $n_F$ is the Fermi-Dirac distribution function.
Since the spin-pumping in the AFI does not affect the Hamiltonian in the superconductor, the distribution function is $f^{s^+}(\varepsilon, \v k) = n_B(\varepsilon, T)$.

\section{Spin current}%
\label{sec:spin_current}
To find the spin current we follow \citet{kato2019} and use that
\begin{align}
  I_s = -\frac{\partial}{\partial t}\left\langle s^z_{\v 0} \right\rangle
  = -i \left\langle \left[H, s^z_{\v 0}\right] \right\rangle.
\end{align}
From the fact that $s^z_{\v 0}$ commutes with $H_\SC + H_\AFI$,
  $\left[s^-_{\v q} , s^z_{\v 0}\right] = s^-_{\v q}$,
and $[A^\dagger,B] = -[A, B^\dagger]^\dagger$,
we find that
\begin{align}
  \left[H, s^z_{\v 0}\right] = 
  \sum_{\v k \in \square} \sum_{\v q \in \Diamond} \left[T^\alpha_{\v q \v k} \alpha_{\v q} s^-_{\v k} + T^{\beta^\dagger}_{\v q \v k} \beta^\dagger_{\v q} s^-_{\v k}
    - \text{h.c.}
  \right].
\end{align}
Thus, the spin current is
\begin{align}
  \label{eq:spinCurr_init}
  I_s(t) =  2
   \sum_{\v k \in \square} \sum_{\v q \in \Diamond}\sum_{\nu\in\{\alpha,\; \beta^\dagger\}}
   \Im\left\langle T^\nu_{\v q \v k} s^-_{\v k}(t)\nu_{\v q}(t) 
   \right\rangle. 
\end{align}

We evaluate this expectation value in the interaction picture and treating the interfacial exchange interaction as a perturbation using the Keldysh formalism.
First, let $G_\psi$ with no superscript denote contour-ordered Green's functions,
\begin{equation}
  G_\psi(\tau_1,\tau_2, \v k) = -i\left\langle\CO \psi_{\v k}(\tau_1) \psi^\dagger_{\v k}(\tau_2)\right\rangle_0,
\end{equation}
where $\CO$ means that $\psi_{\v k}$ and $\psi_{\v k}^\dagger$ are ordered with respect to $\tau_1$ and $\tau_2$ along the complex Keldysh contour, $\mathcal C$.
Next, we define
\begin{align}
  C(\tau_1, \tau_2) \defeq \left\langle \CO T^\nu_{\v q \v k} \nu_{\v q}(\tau_1) s^-_{\v k}(\tau_2)\right\rangle,
\end{align}
where $\nu$ is either $\alpha$ or $\beta^\dagger$.

Going to the interaction picture with $H_\intf$ as the interaction, we get
\begin{multline}
  C(\tau_1, \tau_2) =
  \left\langle \CO T^\nu_{\v q \v k} \nu_{\v q}(\tau_1) s^-_{\v k}(\tau_2)\me{-i\int_{\mathcal C} \dd{\tau} H_\intf(\tau)}\right\rangle_0
  \\
  \approx
  \left\langle \CO\int_{\mathcal C}\dd{\tau} \abs{T^\nu_{\v q \v k}}^2 \nu_{\v q}(\tau_1)\nu^\dagger_{\v q}(\tau)s^+_{-\v k}(\tau) s^-_{\v k}(\tau_2)\right\rangle_0
  \\
  =i\abs{T^\nu_{\v q \v k}}^2 \bigl[G_\nu(\v q) \bullet G_{s^+}(\v k)\bigr](\tau_1, \tau_2),
\end{multline}
where we have used the bullet product $\bullet$ to denote integration of the internal complex time parameter along the Keldysh contour.
In the second equality it was used that
\begin{equation}
  -i\left\langle \CO s^+_{-\v k'}(\tau)s^-_{\v k}(\tau_2)\right\rangle_0
  = \delta_{\v k, \v k'}G_{s^+}(\tau, \tau_2, \v k),
\end{equation}
as can be confirmed by using Wick's theorem.
Next, if we choose $\tau_2$ to be placed later in the contour we have
\begin{equation}
  C(\tau_1, \tau_2) = C^<(\tau_1, \tau_2) = 
  \left\langle T^\nu_{\v q \v k} s^-_{\v k}(\tau_2)\nu_{\v q}(\tau_1)
  \right\rangle.
\end{equation}

From the Langreth rules we have
\begin{align}
  C^<(t,t) = \left[G^R_\nu(\v q) \circ G_{s^+}^<(\v k) + G^<_\nu(\v q) \circ G_{s^+}^A(\v k)\right](t,t),
  \label{eq:c_time}
\end{align}
where the circle product $\circ$ means integration over the internal real time coordinate.
The circle products are the same as normal convolution products, since $G_\psi^R(t_1, t_2)$ and $G_\psi^<(t_1, t_2)$ only depend on time through the relative time $t_1 - t_2$.
Thus, by writing \cref{eq:c_time} in terms of Fourier transformed Green's functions, the circle products become normal products, so, by inserting it into \cref{eq:spinCurr_init},
\begin{multline}
  I_s = 4 \int \frac{\dd{\varepsilon}}{2\pi}
   \sum_{\v k \in \square} \sum_{\v q \in \Diamond}\sum_{\nu\in\{\alpha,\, \beta^\dagger\}}
     \abs{T^\nu_{\v q \v k}}^2\Im G^R_\nu(\varepsilon, \v q) 
     \\\times
     \Im G_{s^+}^R(\varepsilon, \v k)
      \left[f^\nu (\varepsilon, \v q)-f^{s^+}(\varepsilon, \v k)\right],
   \label{eq:spinPumpE}
\end{multline}
where we used that $G_{\psi}^A(\varepsilon) = [G_{\psi}^R(\varepsilon)]^*$.

Inserting \cref{eq:transmissionCoeffs,eq:AFIretarded,eq:AFIdistribution} into \cref{eq:spinPumpE} and using \cref{eq:bogoCoeffs} gives
\begin{equation}
  I_s = I_r + I_U,
\end{equation}
where
\begin{multline}
  \label{eq:regular_curr}
  I_r = -\bar J_A^2 \gamma^2 h_0^2 
  \left(\frac{1}{\left(\Omega - \omega^\alpha_{\v 0}\right)^2 + \left(\eta^\alpha\right)^2} \left[\frac{U_K + (1 - c)}{2 + U_K}\right]^2\right.
  \\
  +
  \left.\frac{1}{\left(\Omega + \omega^\beta_{\v 0}\right)^2 + \left(\eta^\beta\right)^2} \left[\frac{cU_K + (c - 1)}{2 + U_K}\right]^2\right)
  \\ \times
  \sum_{\v k \in \square, l=0}\Im G_{s^+}^R(\Omega, \v k)\delta_{\v k_\parallel, \v 0}
\end{multline}
and
\begin{multline}
  \label{eq:umklapp_curr}
  I_U = -\bar J_A^2 \gamma^2 h_0^2
  \left(\frac{1}{\left(\Omega - \omega^\alpha_{\v 0}\right)^2 + \left(\eta^\alpha\right)^2} \left[\frac{U_K + (1 + c)}{2 + U_K}\right]^2\right.
  \\
  +
  \left.\frac{1}{\left(\Omega + \omega^\beta_{\v 0}\right)^2 + \left(\eta^\beta\right)^2} \left[\frac{c U_K + (c + 1)}{2 + U_K}\right]^2\right)
  \\ \times
  \sum_{\v k\in\square, l=0}
  \Im G_{s^+}^R(\Omega, \v k + \v G)\delta_{\v k_\parallel, \v 0}.
\end{multline}
Here, $U_K = K/(Jz)$ and $c = \bar J_B/\bar J_A$ is the interface asymmetry parameter that gives the degree to which the interface is compensated.
The sums are restricted to include only the $\v k$-vectors that satisfy $\delta_{\v k_\parallel, \v 0}=1$ with $l=0$ and $\v G$ is the vector that connects these to the $\v k$-vectors with $l=1$.
When both the SC and AFI are cubical with a lattice parameter $a$ and a compensated interface, then $\v G = \pi (\uv x + \uv y + \uv z)/a$.
In order for the Umklapp scattering to produce a nonzero $I_U$, it is necessary that there exists $\v k, \v q \in \square$ such that both $\v q$ and $\v q + \v k + \v G$ are close to the Fermi surface and $\delta_{\v k_\parallel, \v 0} = 1$.
In a cubical lattice the minimal value of $\v k + \v G$ is $\sqrt{2}\pi/a$, so the maximal diameter of the Fermi surface must be at least $\sqrt{2}\pi/a$.
The Umklapp current is also zero if the interface is fully uncompensated.
In this case there is no Umklapp scattering and the current is simply $I_s = I_r$ with $c = 0$.

\section{Numerical results}
\begin{figure}[htpb]
  \centering
  \includegraphics[width=1.0\linewidth]{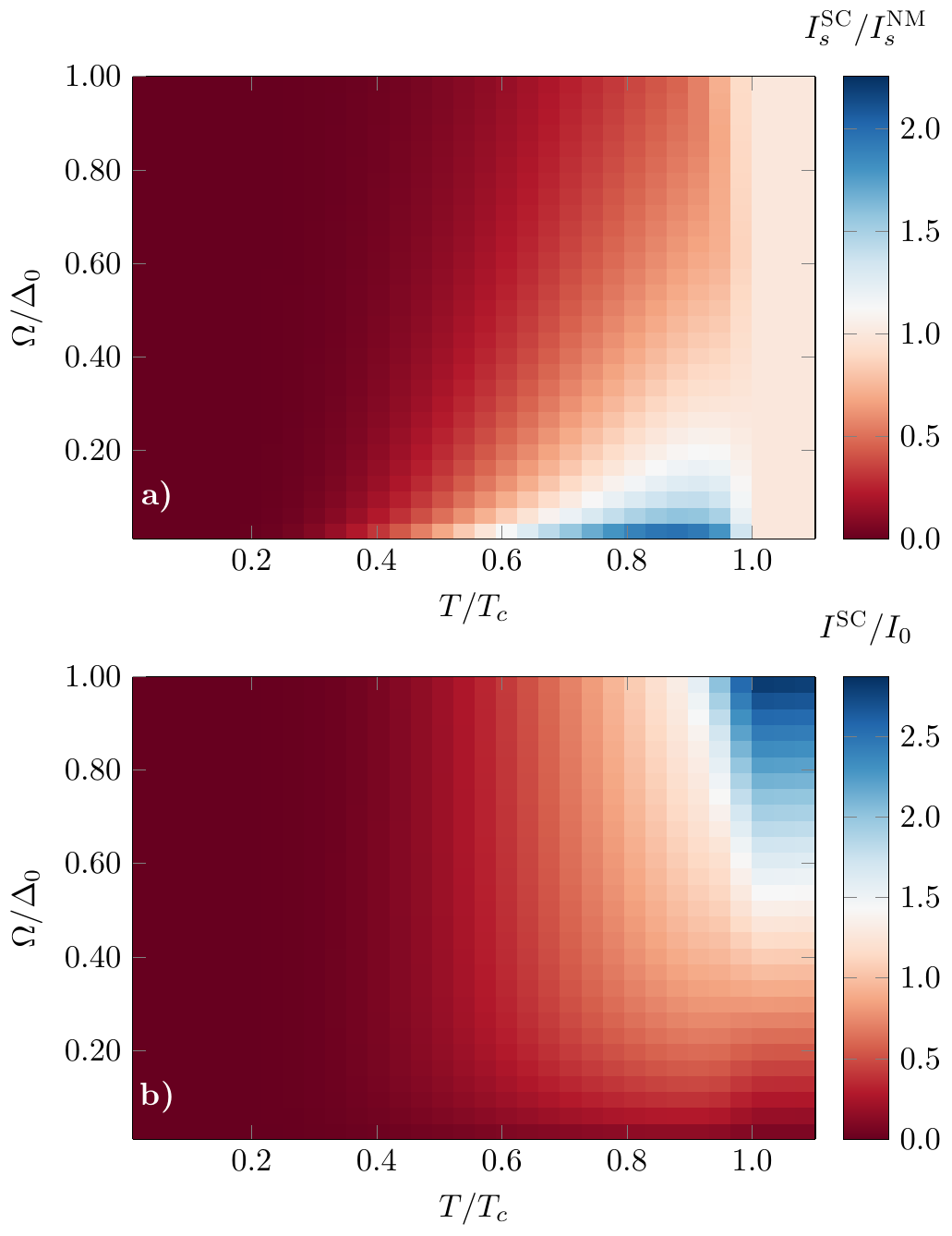}
  \caption{The spin-current into a superconductor with gap given by \cref{eq:gap}, $I^\SC_s$, for different precession frequencies $\Omega$ and temperatures $T$ and normalized by the normal state spin-current, $I^\NM_s$, found by setting $\Delta = 0$ in a) and the constant $I_0 = \gamma^2 h_0^2 \bar{J}_A^2 N_SN_S^\perp/\left[(2\pi)^4 \Delta_0\right]$ in b).
  $N_S^\perp$ is the number of lattice points in the superconductor in the direction transverse to the interface,
$\Delta_0$ is the superconducting gap at $T=0$ and $T_c$ is the critical temperature.}%
  \label{fig:2d}
\end{figure}

Next we show numerical results for a cubical lattice with lattice constant $a$ such that
\begin{equation}
  \xi_{\v k} = -2t\sum_{i\in\{x, y, z\}} \cos(a k_i) - \mu,
\end{equation}
where $t$ is the hopping parameter.
In \cref{fig:2d} we show the spin current into the superconductor, $I^\SC_s$, for different temperatures $T$ and precession frequencies $\Omega$, normalized by the normal state value, $I^\NM_s$ in \cref{fig:2d}~a) and a constant in \cref{fig:2d}~b).
In this case we have used $\mu = -4t$, which means that $I_U = 0$.
However, we find that both $I_r$ and $I_U$ scale in the same way as functions of $\Omega$ and $T$ also for other values of $\mu$.
In \cref{fig:2d} we have also used $U_k = 0.01$, which is close to the reported value for MnF$_2$~\cite{hagiwara1996,johnson1959}, $t = 1000\Delta_0$, $c = 0.5$, $\eta^\alpha = \eta^\beta = \Delta_0 \times 10^{-4}$ and $\omega_{\v 0}^\alpha = \omega_{\v 0}^\beta = 4\Delta_0$.
This corresponds to a resonance frequency of $\SI{1}{\tera\hertz}$ when $\Delta_0 = \SI{1}{\milli\electronvolt}$.
From \cref{fig:2d}~b) we see that the normal state spin current at $T > T_c$ scales linearly with $\Omega$ as expected.
In comparison, the spin current changes only very slowly with $\Omega$ in the superconducting state.
This is consistent with the physical picture that it is the availability of quasiparticles rather than unoccupied states that limits the current in the superconducting state.

\begin{figure}[htpb]
  \centering
  \includegraphics[width=1.0\linewidth]{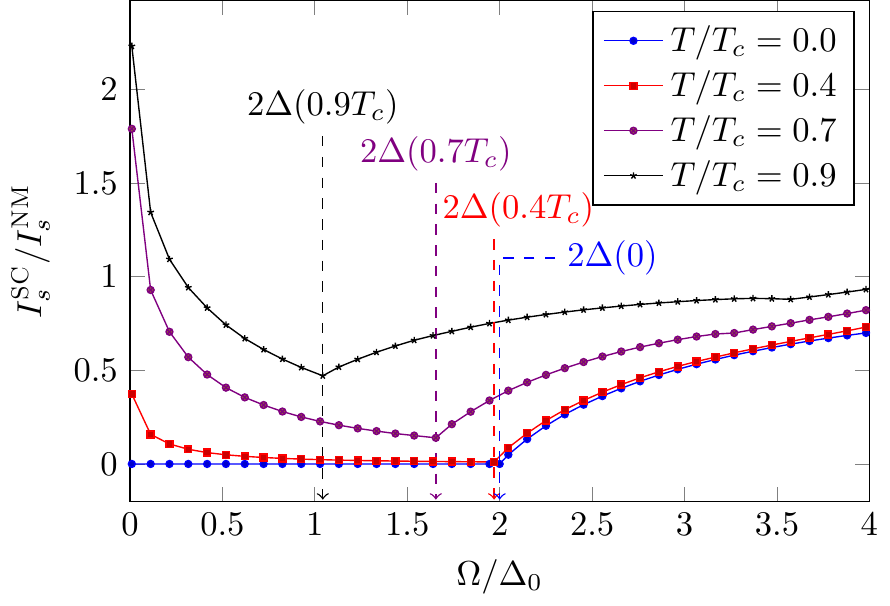}
  \caption{The spin-current into a superconductor, $I^\SC_s$ normalized by the normal state spin-current, $I^\NM_s$, found by setting the gap $\Delta = 0$, as a function of the spin-pumping precession frequency $\Omega$. Here, $\Delta(T)$ is the energy gap that solves \cref{eq:gap} and $T_c$ is the critical temperature.}%
  \label{fig:slices}
\end{figure}

As one can see from \cref{fig:2d}~a), the spin-current in the superconductor is peaked at small frequencies close to the critical temperature, where it can be more than twice as large as the normal-metal spin-current.
This is similar to the results for spin-currents in superconductor-ferromagnetic bilayers~\cite{kato2019,inou2017}.
\Cref{fig:slices} shows the ratio $I_s^\SC/I_s^\NM$ as a function of $\Omega/\Delta_0$ for various $T$.
It can be seen that at zero temperature the spin current in the superconducting case is zero for $\Omega < 2\Delta_0$.
For $T > 0$ the ratio $I_s^\SC/I_s^\NM$ initially decreases as $\Omega$ increases and reaches a minimum at $\Omega = 2 \Delta(T)$.

This can be understood physically in the following way.
The spin-current is generated by spin-flip scatterings which excite particles by energy $\Omega$ and flip their spin.
This can be seen from \cref{eq:Hint,eq:BCSspinSusc,eq:regular_curr,eq:umklapp_curr} when $\eta^\SC \ll 1$.
In this case the sum in \cref{eq:BCSspinSusc} only contribute to imaginary part of $G^R_{s^+}(\Omega, \v k)$ when $\tilde\omega - \omega = \Omega$, and only when $n_F(\omega,T)-n_F(\tilde\omega,T)\neq 0$.
In the normal metal case there is a number of electrons proportional to $\Omega$ around the Fermi surface which can be excited to an available state.
Hence, the dynamic spin-susceptibility is proportional to $\Omega$.

In a superconductor the spin-flip scatterings can happen by breaking a Cooper pair or exciting a quasiparticle from above the gap to a higher energy.
When $\Omega < 2\Delta(T)$ only the latter is possible.
Thus, in order to get a nonzero spin-current when $\Omega < 2\Delta(T)$ the temperature must be large enough for quasiparticle states above the gap to be occupied.
This is why, in \cref{fig:slices}, the current is identically zero in the superconductor when $T = 0$ and $\Omega < 2\Delta_0$.
On the other hand, when the temperature is close to the critical temperature there can be many available quasiparticles available because the density of states is peaked around the gap.
This peak in the density of states is why the spin-current in a superconductor can be larger than the spin-current in a normal metal, but only when the temperature is close to the critical temperature.
It is also only larger when $\Omega \ll \Delta(T)$, which is because the lack of states below the gap in the superconductor means that the spin susceptibility can not increase as fast as in the normal state when $\Omega$ increases.
In the normal state there is a range of energies $\propto\Omega$ around the Fermi surface that can be excited to an available state, but in the superconducting state the number of states that can be excited is limited by the number of quasiparticles present.
Increasing $\Omega$ therefore decreases the ratio $I_s^\SC/I_s^\NM$ when $\Omega < 2 \Delta(T)$, as can be seen in \cref{fig:2d,fig:slices}.
At $\Omega = 2 \Delta(T)$ the breaking of Cooper pairs becomes possible as a spin-transfer mechanism, which is why $I_s^\SC/I_s^\NM$ starts to increase.
This can be seen most clearly in \cref{fig:slices}.

\begin{figure}[htpb]
  \centering
  \includegraphics[width=1.0\linewidth]{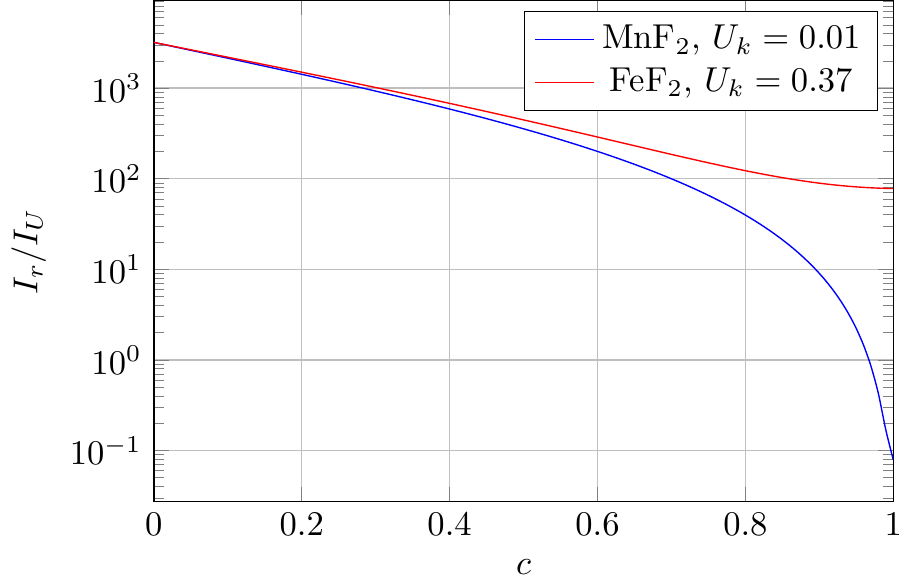}
  \caption{The ratio of the spin-current contribution from the regular scattering channel, $I_r$, and the Umklapp scattering channel, $I_U$, as a function of the interface asymmetry parameter $c$ for $\Omega/\Delta_0 = 0.1$, $T/T_c = 0.9$, $\mu = -2.2t$.
  The results are shown for easy axis anisotropy values $U_k = 0.01$ and $U_k = 0.37$, where the former is found in MnF$_2$ and the latter is found in FeF$_2$~\cite{hagiwara1996,johnson1959,ohlman1961}.}%
  \label{fig:umklapp}
\end{figure}

\Cref{fig:umklapp} shows the ratio between regular spin-current $I_r$ and the Umklapp contribution $I_U$ for $\mu = -2.2t$.
The result is shown for axis anisotropy $U_k = 0.01$, which correspond to MnF$_2$~\cite{hagiwara1996,johnson1959}, and $U_k = 0.37$, corresponding to FeF$_2$~\cite{ohlman1961}.
In both cases the regular current dominates when the interface asymmetry parameter $c$ is small, meaning that the superconductor is coupled more strongly to one of the sublattices in the AFI.
The Umklapp contribution becomes more important as $c$ increases and when $U_k$ is small the Umklapp contribution eventually becomes larger than the contribution from the regular scattering channel.
This is consistent with the work by \citeauthor{kamra2017} showing that the in the absence of easy-axis anisotropy the cross-sublattice contribution quench the spin-current from the regular scattering channel~\cite{kamra2017}.
However, here we see that if we include the Umklapp scattering the spin-current will not go all the way to zero, even in the absence of easy-axis anisotropy.
Mathematically, this can be seen from \cref{eq:regular_curr,eq:umklapp_curr}: when $U_k = 0$ we have $I_r \propto (1-c)^2$ and $I_U \propto (1+c)^2$.
However, when $U_k= 0.37$ the regular contribution remains dominant for all values of $c$.

\section{Experimental detection}
\label{sec:experiment}
Although the spin-current can be enhanced in SC/AFI bilayers as compared to NM/AFI bilayers, it can be difficult to observe this enhancement experimentally.
This is because the spin-current is strongly peaked around the antiferromagnetic resonance frequencies $\omega_{\v 0}^{\alpha/\beta}$.
In antiferromagnets this resonance frequency is on the order of $\SI{1}{\tera\hertz}$, which is much larger than in ferromagnets~\cite{baltz2018}.
This is an advantage for spintronics as it allows for ultrafast information processing, but in the context of this paper it means that observation of the enhancement produced by the superconducting order is hard to experimentally verify.
A resonance frequency of $\SI{1}{\tera\hertz}$ means that the spin-current is most easily observed at $\Omega/\Delta_0 \approx 4$, assuming that $\Delta_0 = \SI{1}{\milli\electronvolt}$, but form \cref{fig:2d} we see that the current is enhanced only for $\Omega/\Delta_0 < 0.2$.

In order to observe the strong suppression of spin-current at low temperatures, it is necessary to probe frequencies below $2\Delta_0$.
This is also below $\SI{1}{\tera\hertz}$, but not out of reach.
The resonance frequency of MnF$_2$, which was used in the detection of spin-pumping by \citeauthor{vaidya2020}, was reported to be around $\SI{250}{\giga\hertz}$~\cite{vaidya2020}.
This corresponds to $\Omega \approx \SI{1}{\milli\electronvolt} \approx \Delta_0$, which makes the low-temperature suppression shown in \cref{fig:2d,fig:slices} detectable.

\begin{figure}[htpb]
  \centering
  \includegraphics[width=1.0\linewidth]{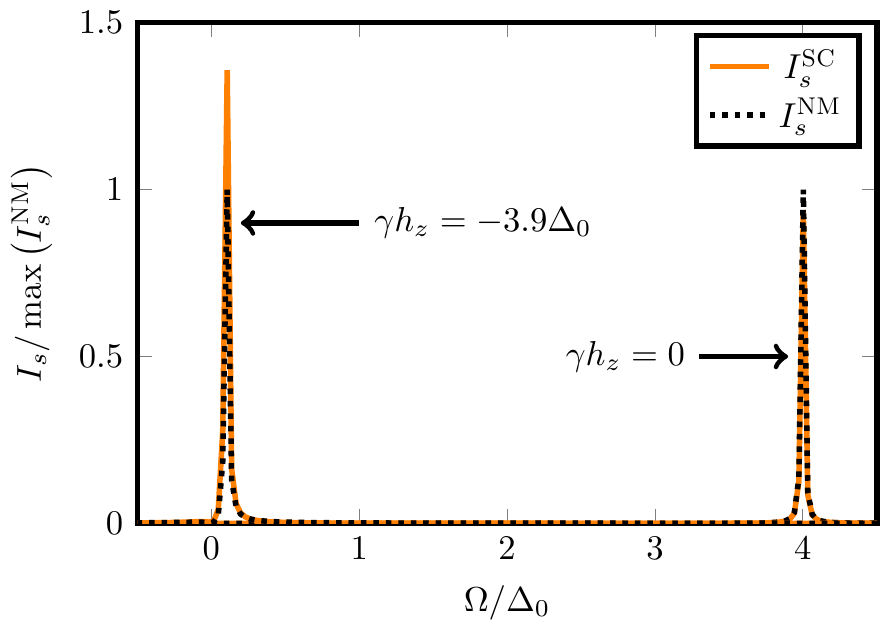}
  \caption{The superconductor spin-current $I_s^\SC$ and normal metal spin-current $I_s^\NM$ normalized by the maximal value of $I_s^\NM$ as a function of the precession frequency $\Omega$ for two different values of constant external magnetic field $h_z$. Here, $\gamma$ is the gyromagnetic ratio, $T/T_c = 0.9$, $\eta^{\alpha/\beta}/\Delta_0 = 0.01$, $c=0.5$, $U_k = 0.01$, $T_c$ is the critical temperature and $\Delta_0$ is the superconducting gap at zero temperature.}%
  \label{fig:peaks}
\end{figure}

One way to potentially detect the spin-current enhancement at low frequencies is to apply a constant magnetic field along the $z$-axis.
This was also done by \citeauthor{vaidya2020}, who reduced the frequency of MnF$_2$ to $\SI{120}{\giga\hertz}$ by applying a magnetic field of $\SI{4.7}{\tesla}$.
From \cref{eq:bogoCoeffs,eq:regular_curr,eq:umklapp_curr} we see that the resonance frequencies are
\begin{equation}
  \omega_\text{res} = \omega_0\sqrt{U_k(2+U_k)} \pm \gamma h_z.
\end{equation}
where $\omega_0 = JzS$.
Thus, by applying a magnetic field of $\omega_0\sqrt{U_k(2+U_k)}/\gamma$, the resonance frequency can be pushed well below $\Delta_0$, making the enhancement in spin-current due to superconductivity detectable.
This is illustrated in \cref{fig:peaks}.
At $\gamma h_z = 0$ the peak is at $\Omega = 4\Delta_0$, where the peak in $I_s^\SC$ is only slightly smaller than the peak in $I_s^\NM$, in accordance with \cref{fig:slices}.
However, when $\gamma h_z = -3.9\Delta_0$, the peak is shifted to $\Omega = 0.1\Delta_0$ and the peak in the superconducting case is taller.
How large the applied magnetic field is required to be depend on the gyromagnetic ratio $\gamma$ as well as $\omega_0$ and $U_k$, but it will in general be several tesla.
Experimental ingenuity is therefore required in order to make sure that the superconductivity is not completely suppressed by the magnetic field.
This could for instance be done by shielding the superconductors or using superconductors that can withstand large magnetic field from a certain direction, such as Ising superconductors.

\section{Conclusion}%
\label{sec:conclusion}
We have derived an expression for the spin-current in SC/AFI bilayers undergoing spin-pumping, valid for both compensated and uncompensated interfaces and taking into consideration both the regular scattering channel and the Umklapp scattering channel.
We found that for temperature $T$ well below the critical temperature $T_c$, the spin-current is strongly suppressed as long as the precession frequency of the applied magnetic field is less than $2\Delta(T)$.
This is because the energy gap in the superconductor inhibits spin-flip scatterings below the gap and there are few quasiparticles present that can be scattered to higher energies.
However, at temperatures close to $T_c$ there are quasiparticles present and because of their large density of states close to the gap, the spin-current can be more than twice as large as for NM/AFI bilayers when the precession frequency is significantly less than the gap.
The spin-current contribution from the Umklapp channel is typically much smaller than the contribution from the regular scattering channel, but it can be significant if the Fermi surface is large, the easy axis anisotropy is small and the interface is compensated.

The relevant precession frequencies where the spin-current in SC/AFI is enhanced compared to NM/AFI is much lower than the typical resonance frequencies of antiferromagnets, which makes the detection of this effect experimentally challenging.
A possible solution lies in the shifting of the resonance frequency by a static magnetic field.

\begin{acknowledgments}
This work was supported by the Research Council of Norway through grant 240806, and its Centres of Excellence funding scheme grant 262633 ``\emph{QuSpin}''. J. L. also acknowledge funding from the NV-faculty at the Norwegian University of Science and Technology. 
\end{acknowledgments}

\appendix
\section{AFI Green's functions}%
\label{sec:afi_green_s_functions}
In this section we calculate the correction to the magnon Green's functions due to the precessing external magnetic field.
The Hamiltonian for the antiferromagnetic insulator is given by \cref{eq:H_afi}, and we treat 
\begin{align}
  V \defeq \sqrt{2N_AS} (u_{\v 0} + v_{\v 0}) \gamma\left[h^-\left(\alpha_{\v 0} + \beta^\dagger_{\v 0}\right) + h^+\left(\alpha^\dagger_{\v 0} + \beta_{\v 0}\right)\right]
  \label{eq:V}
\end{align}
as a perturbation.
In order to calculate $G_{\nu}^R$ and $G_{\nu}^<$, where $\nu\in\{\alpha, \beta^\dagger\}$, we will first calculate the contour-ordered Green's function.
This, in turn, is done by adding an infinitesimal imaginary part to the otherwise real time coordinates and integrating over the complex Keldysh contour.

To second order in $V$, the contour-ordered Green's function is
\begin{multline} 
  G_\nu(\tau_1,\tau_2, \v k) = 
  -i\left\langle \CO \nu_{\v k}(\tau_1)\nu^\dagger_{\v k}(\tau_2) \me{-i\int_\mathcal{C} \dd{\tau} V(\tau)} \right\rangle_0
  \\
  =-i\left\langle \CO \nu_{\v k}(\tau_1)\nu^\dagger_{\v k}(\tau_2) \right\rangle_0 
  - \left\langle \CO \int_\mathcal{C} \dd{\tau} \nu_{\v k}(\tau_1)\nu^\dagger_{\v k}(\tau_2) V(\tau) \right\rangle_0 
  \\
  +i\left\langle \CO \int_\mathcal{C} \dd{\tau'} \dd{\tau} \nu_{\v k}(\tau_1)\nu^\dagger_{\v k}(\tau_2) V(\tau) V(\tau') \right\rangle_0 
  + \mathcal O(V^3),
\end{multline}
where $\CO$ means ordering along the Keldysh contour $\mathcal{C}$ and the subscript 0 means that the expectation values are evaluated in the absence of $V$.
The first order term is odd in magnon operators and is therefore zero.
Inserting \cref{eq:V}, the correction to the equilibrium Green's function is
\begin{multline}
  \Delta G_\nu(t_1, t_2, \v k) \defeq G_\nu(t_1,t_2, \v k)  - G^0_\nu(t_1,t_2, \v k) 
  \\
  = i\lambda^2
  \left\langle \CO \int_\C \dd{\tau'} \dd{\tau} \nu_{\v k}(t_1)\nu^\dagger_{\v 0}(\tau) h^+(\tau)h^-(\tau')\nu_{\v 0}(\tau') \nu^\dagger_{\v k}(t_2) \right\rangle_0,
\end{multline}
where
\begin{equation}
  \lambda = \sqrt{2N_AS} (u_{\v 0} + v_{\v 0}) \gamma.
\end{equation}
We can use Wick's theorem to evaluate the rewrite this as
\begin{multline}
  \left\langle \CO \int_\C \dd{\tau'} \dd{\tau} \nu_{\v k}(t_1)\nu^\dagger_{\v 0}(\tau) h^+(\tau)h^-(\tau')\nu_{\v 0}(\tau') \nu^\dagger_{\v k}(t_2) \right\rangle_0
  \\
  = 
  \int_C \dd{\tau'} \dd{\tau}h^+(\tau)h^-(\tau') 
  \Biggl[
    \left\langle \CO \nu_{\v k}(t_1)\nu^\dagger_{\v 0}(\tau)\right\rangle_0
    \left\langle \CO\nu_{\v 0}(\tau') \nu^\dagger_{\v k}(t_2) \right\rangle_0
    \\
    +
    \left\langle \CO \nu_{\v k}(t_1) \nu^\dagger_{\v k}(t_2)\right\rangle_0
    \left\langle \CO\nu_{\v 0}(\tau') \nu^\dagger_{\v 0}(\tau)\right\rangle_0
  \Biggr].
\end{multline}
The second term is zero, which we show in the following.
First, define
\begin{equation}
  \Sigma(\tau_1, \tau_2) = h^+(t_1)h^-(t_2) = \left\langle \CO h^+(\tau_1)h^-(\tau_2)\right\rangle.
\end{equation}
Then,
\begin{multline}
  \int_\C \dd{\tau'} \dd{\tau}h^+(\tau)h^-(\tau') 
    \left\langle \CO \nu_{\v k}(\tau_1) \nu^\dagger_{\v k}(\tau_2)\right\rangle_0
    \left\langle \CO\nu_{\v 0}(\tau') \nu^\dagger_{\v 0}(\tau)\right\rangle_0
    \\
    = -G_\nu^0(\tau_1, \tau_2, \v k) \int_\C \dd{\tau} 
    \left[\Sigma \bullet G_\nu^0\right]_{\v k = \v 0}(\tau, \tau) \\
    =-G_\nu^0(\tau_1, \tau_2, \v k) 
    \left(\int_{-\infty}^\infty \dd{t} + \int_{\infty}^{-\infty} \dd{t}\right)
    \left[\Sigma \bullet G_\nu^0\right]_{\v k = \v 0}(t, t)
    = 0,
\end{multline}
where it was used that $\C$ goes from $-\infty - i\delta$ to $\infty - i\delta$ and then from $\infty + i\delta$ to $-\infty + i\delta$ with $\delta \in \mathbb{R}$ being an infinitesimal.
The bullet product is 
\begin{equation}
  (A\bullet B)(\tau_1, \tau_2) = \int_\C \dd{\tau} A(\tau_1, \tau)B(\tau, \tau_2).
\end{equation} 
Hence, we are left with
\begin{align}
  \Delta G_\nu(\tau_1,\tau_2, \v k) = -i\delta_{\v k, \v 0} \lambda^2
  \left(G_\nu^0\bullet\Sigma\bullet G_\nu^0\right)(\tau_1, \tau_2).
\end{align}

To get the real-time Green's functions we can use the Langreth rules.
If
\begin{subequations}
  \begin{align}
    C(\tau_1, \tau_2) &= (A\bullet B)(\tau_1, \tau_2), \\
    D(\tau_1, \tau_2) &= (A\bullet B \bullet C)(\tau_1, \tau_2),
  \end{align}
\end{subequations}
where $A$ and $B$ are contour-ordered functions,
then the corresponding advanced, retarded and lesser Green's functions satisfy~\cite{rammer2007}
\begin{subequations}
  \label{eq:rammerIds}
  \begin{align}
    \label{eq:rammerIdsTwo}
    C^< &= A^R\circ B^< + A^<\circ B^A, \\
    \label{eq:rammerIdsTwoR}
    C^{R/A} &= A^{R/A}\circ B^{R/A}, \\
    \label{eq:rammerIdsThree}
    D^< &= A^R\circ B^R\circ C^< + A^R\circ B^<\circ C^A\nonumber \\ 
        &\phantom{=}+ A^<\circ B^A\circ C^A,\\
    \label{eq:rammerIdsThreeR}
    D^{R/A} &= A^{R/A}\circ B^{R/A}\circ C^{R/A},
  \end{align}
\end{subequations}
where the circle product is
\begin{equation}
  (A\circ B)(t_1, t_2) = \int_{-\infty}^{\infty} \dd{t} A(t_1, t)B(t, t_2).
\end{equation} 

Using \cref{eq:rammerIdsThree} as well as $\Sigma^< = \Sigma$ and $\Sigma^R = \Sigma^A = 0$ we see that $\Delta G_\nu^{R/A} = 0$ and
\begin{align}
  \Delta G^<_\nu(t_1,t_2, \v k) = -i\delta_{\v k, \v 0} \lambda^2
  \left(G_\nu^{R}\circ\Sigma\circ G_\nu^{A}\right)(t_1, t_2).
  \label{eq:dgless}
\end{align}
Next, if we let $h_x(t)=h_0 \cos(\Omega t)$ and $h_y(t)=-h_0 \sin(\Omega t)$ we get
$h^\pm(t) = h_0\exp(\mp i\Omega t)$, so
\begin{align}
  \Sigma(t_1, t_2) = h_0^2\me{-i\Omega(t_1-t_2)}.
\end{align}

The circle products in \cref{eq:dgless} reduce to normal convolutions because $G^0_\nu$ and $\Sigma$ only depend on the relative time.
Thus, they further reduce to ordinary products in energy-space.
The Fourier transform of $\Sigma$ is
\begin{equation}
  \Sigma(\varepsilon) = \int_{-\infty}^\infty \dd{(t_1 - t_2)} \Sigma(t_1, t_2) \me{i\varepsilon (t_1 - t_2)} = 2\pi h_0^2 \delta(\varepsilon - \Omega).
\end{equation}
We also have that~\cite{rammer2007}
\begin{equation}
  G_\nu^{A}(\varepsilon, \v k) = 
\left[G_\nu^{R}(\varepsilon, \v k)\right]^*,
\end{equation}
so, to second order in $h$,
\begin{align}
  \Delta G^<_\nu(\varepsilon, \v k) = -2i\pi h_0^2 \lambda^2 \abs{G_\nu^{R}(\varepsilon, \v k)}^2 \delta_{\v k, \v 0} \delta(\varepsilon - \Omega).
\end{align}
Inserting this into the definition of the distribution function and using \cref{eq:AFIretarded} finally gives us \cref{eq:AFIdistribution}.

\section{BCS dynamic spin susceptibility}%
\label{sec:bcs_dynamic_spin_susceptibility}
To calculate $\Im G_{s^+}^R(\varepsilon, \v k)$ we will use the imaginary time Green's function~\cite{bruus2004}
\begin{equation}
  \bar G_{s^+}(\tau_1, \tau_2, \v k) = -\langle\TO_\tau s^+_{-\v k}(\tau_1) s^-_{\v k}(\tau_2)\rangle,
\end{equation}
where $\TO_\tau$ means time-ordering in $\tau$,
together with the connection through analytical continuation,
\begin{equation}
   G_{s^+}^R(\varepsilon, \v k) = \bar G_{s^+}(\varepsilon + i\eta^\SC, \v k),
   \label{eq:anCont}
\end{equation}
where
\begin{equation}
  \label{eq:chiMats}
  \bar G_{s^+}(i\omega_n, \v k) = \int_0^\beta \dd{(\tau_1 - \tau_2)} 
  \bar G_{s^+}(\tau_1, \tau_2, \v k) \me{i \omega_n (\tau_1 - \tau_2)}
\end{equation}
and 
\begin{equation}
  \omega_n = \frac{2n\pi}{\beta}
\end{equation}
are bosonic Matsubara frequencies.
The inverse temperature is $\beta = 1/T$.

We will also make use of the Nambu spinors
\begin{equation}
  \phi_{\v k}^\dagger = \mqty(c_{\v k,\up}^\dagger & c_{-\v k,\dn}).
\end{equation}
With these spinors we can write
\begin{align}
  &s_{\v k}^- = \sum_{\v q} \phi_{-\v q, 2} \phi_{\v q + \v k, 1},
  &&s_{-\v k}^+ = \sum_{\v q} \phi^\dagger_{\v q + \v k, 1} \phi^\dagger_{-\v q, 2}.
\end{align}
Thus,
\begin{multline}
  \bar G_{s^+}(\tau_1, \tau_2, \v k) \\
  = -\sum_{\v q \v q'}\left\langle\TO_\tau \phi^\dagger_{\v q + \v k, 1}(\tau_1) \phi^\dagger_{-\v q, 2}(\tau_1) \phi_{-\v q', 2}(\tau_2) \phi_{\v q' + \v k, 1}(\tau_2)\right\rangle \\
  = \sum_{\v q \v q'}\Biggl(\left\langle\TO_\tau \phi^\dagger_{\v q + \v k, 1}(\tau_1) \phi_{-\v q', 2}(\tau_2) \right\rangle\left\langle\TO_\tau\phi^\dagger_{-\v q, 2}(\tau_1) \phi_{\v q' + \v k, 1}(\tau_2)\right\rangle\\
  - \left\langle\TO_\tau \phi^\dagger_{\v q + \v k, 1}(\tau_1) \phi_{\v q' + \v k, 1}(\tau_2) \right\rangle\left\langle\TO_\tau \phi^\dagger_{-\v q, 2}(\tau_1) \phi_{-\v q', 2}(\tau_2)\right\rangle\Biggr)\\
  = \sum_{\v q}\Bigl[\mathcal G_{1,2}(\tau_2, \tau_1, \v q + \v k)\mathcal G_{2,1}(\tau_2, \tau_1, -\v q) \\
    - \mathcal G_{1,1}(\tau_2, \tau_1, \v q + \v k)\mathcal G_{2,2}(\tau_2, \tau_1, -\v q)\Bigr],
\end{multline}
where
\begin{multline}
  \mathcal G(\tau_1, \tau_2, \v k) = -\left\langle\TO_\tau \phi_{\v k}(\tau_1) \phi^\dagger_{\v k}(\tau_2) \right\rangle
  \\
  = \frac 1 \beta \sum_n \frac{1}{(i\nu_n)^2 - \xi_{\v k}^2 - \abs{\Delta}^2}
  \mqty(i\nu_n + \xi_{\v k} & -\Delta \\ -\Delta^* & i\nu_n - \xi_{\v k}) \me{-i\nu_n(\tau_1-\tau_2)},
  \label{eq:greenBCS}
\end{multline}
is the BCS single-particle Green's function.
Here, $\nu_n = (2n+1)\pi/\beta$ are fermionic Matsubara frequencies.
Inserting this into \cref{eq:chiMats}, we get
\begin{multline}
  \bar G_{s^+}(i\omega_n, \v k)
  = T \sum_{\v q, m}
  \Bigl[\mathcal G_{1,2}(-i\nu_m-i\omega_n, \v q + \v k)\mathcal G_{2,1}(i\nu_m, -\v q)\\
  - \mathcal G_{1,1}(-i\nu_m-i\omega_n, \v q + \v k)\mathcal G_{2,2}(i\nu_m, -\v q)\Bigr] \\
  = T \sum_{\v q, m}
  \Bigl[\mathcal G_{1,2}(i\nu_m + i\omega_n, \v q + \v k)\mathcal G_{2,1}(i\nu_m, -\v q)\\
  + \mathcal G_{2,2}(i\nu_m + i\omega_n, \v q + \v k)\mathcal G_{2,2}(i\nu_m, -\v q)\Bigr] \\
  = \frac{1}{2\beta} \sum_{\v q, m}
  \Tr\left[\mathcal G(i\nu_m + i\omega_n, \v q + \v k)\mathcal G(i\nu_m, \v q)\right].
  \label{eq:chi1}
\end{multline}
In the last equality we have used that $\mathcal G(i\nu_m, -\v q) = \mathcal G(i\nu_m, \v q)$, $\mathcal G_{1,2}(i\nu_n, \v k)\mathcal G_{2,1}(i\nu_m, \v q) = \mathcal G_{2,1}(i\nu_n, \v k)\mathcal G_{1,2}(i\nu_m, \v q)$ and
\begin{multline}
  \sum_{\v q, m}\mathcal G_{1,1}(i\nu_m + i\omega_n, \v q + \v k)\mathcal G_{1,1}(i\nu_m, \v q)
  \\
  = 
  \sum_{\v q', k}\mathcal G_{1,1}(-i\nu_k, \v q')\mathcal G_{1,1}(-i\nu_k-i\omega_n, \v q' + \v k) \\
  = \sum_{\v q', k}\mathcal G_{2,2}(i\nu_k, \v q')\mathcal G_{2,2}(i\nu_k + i\omega_n, \v q' + \v k),
\end{multline}
In the first equality, we introduced $\v q' = -\v q - \v k$ and $i\nu_k = -i\nu_m - i\omega_n$, and in the last equality we used that $\mathcal G_{2,2}(i\nu_k, \v q') = -\mathcal G_{1,1}(-i\nu_k, \v q')$.

Next, we can use the spectral form,
\begin{equation}
  \mathcal G(i\nu_m, \v q)
  =\int_{-\infty}^\infty \frac{\dd\omega}{(-\pi)}
  \frac{\Im\mathcal G(\omega + i\eta^\SC, \v q)}{i\nu_m - \omega},
\end{equation}
and the Matsubara sum identity
\begin{equation}
  \frac 1 \beta \sum_m \frac{1}{i\nu_m + i\omega_n - \tilde\omega}\times\frac{1}{i\nu_m - \omega} =
  \frac{n_F(\omega, T)- n_F(\tilde\omega, T)}{i\omega_n - (\tilde\omega - \omega)},
\end{equation}
where we have used that $\nu_m$ are fermionic Matsubara frequencies, giving rise to the Fermi-Dirac distribution function $n_F$.
We have also used that $n_F(\omega - i\omega_n) = n_F(\omega)$ since $\omega_n$ is a bosonic Matsubara frequency.
Additionally, \cref{eq:greenBCS} gives, assuming $\Delta$ real,
\begin{multline}
  \Im\mathcal G(\omega + i\eta^\SC, \v k) = -\frac{\pi}{2\sqrt{\xi_{\v k}^2 + \abs{\Delta}^2}}
  \mqty(\omega + \xi_{\v k} & -\Delta \\ -\Delta & \omega - \xi_{\v k})
  \\ \times
  \left[\delta\left(\omega - \sqrt{\xi_{\v k}^2 + \abs{\Delta}^2}\right) - \delta\left(\omega +\sqrt{\xi_{\v k}^2 + \abs{\Delta}^2}\right)\right]
\end{multline}
in the limit $\eta^\SC \to 0^+$.
Hence, if we define $E_{\v k} \defeq \sqrt{\xi_{\v k}^2 + \abs{\Delta}^2}$,
\begin{multline}
  \lim_{\eta^\SC \to 0^+}\Tr\left[\Im\mathcal G(\tilde\omega + i\eta^\SC, \v q + \v k)\Im\mathcal G(\omega + i\eta^\SC, \v q)\right] \\
  = \pi^2 \frac{\omega\tilde\omega + \xi\tilde\xi + \Delta^2}{2E\tilde E}
  \left[\delta\left(\omega - E\right) - \delta\left(\omega + E\right)\right]
  \\ \times
  \left[\delta\left(\tilde\omega - \tilde E\right) - \delta\left(\tilde\omega + \tilde E\right)\right],
\end{multline}
where $\xi = \xi_{\v q}$, $\tilde \xi = \xi_{\v q + \v k}$, $E = E_{\v q}$ and $\tilde E = E_{\v q + \v k}$.
Inserting this into \cref{eq:chi1} gives
\begin{multline}
  \bar G_{s^+}(i\omega_n, \v k) = -\frac 1 4 \sum_{\v q}\sum_{\omega = \pm E}\sum_{\tilde\omega = \pm\tilde E} \frac{\omega\tilde\omega + \xi\tilde\xi + \Delta^2}{\omega\tilde\omega}
  \\ \times
  \frac{n_F(\tilde\omega, T_\SC)- n_F(\omega, T_\SC)}{i\omega_n - (\tilde\omega - \omega)}.
\end{multline}
From \cref{eq:anCont} we then finally have \cref{eq:BCSspinSusc}.

\bibliography{bibliography}

\end{document}